\documentclass[ reprint, twocolumn, nofootinbib, showkeys, aps, prd]{revtex4-2}
\usepackage{amsmath}
\usepackage{amssymb}
\usepackage[export]{adjustbox}
\usepackage{fix-cm}
\usepackage{adjustbox}
\usepackage{latexsym,amsmath,amssymb}
\RequirePackage{graphicx}
\RequirePackage{mathptmx}
\usepackage[arrowdel]{physics}
\usepackage{graphicx,epstopdf}
\usepackage{latexsym}
\usepackage{slashed}
\usepackage{bm}
\usepackage{cases}
\usepackage{multirow}
\usepackage{color}
\usepackage{bm}
\usepackage{dcolumn}
\usepackage{subfigure}
\usepackage{subcaption}
\usepackage{lipsum}
\usepackage{diagbox}

\usepackage[colorlinks = true, linkcolor = blue, urlcolor  = blue, citecolor = red, anchorcolor = blue]{hyperref}

\begin{document}

\preprint{APS/123-QED}

\title{Primordial black holes  from a type-IIB/F theory  inflaton potential}
\author{Chitrak Sarkar}
\author{Amitava Choudhuri}
 \email{amitava\_ch26@yahoo.com}
 \author{Buddhadeb Ghosh}
%\email{bg@gmail.com}
\affiliation{Department of Physics, The University of Burdwan, Golapbag, Purba Bardhaman, West Bengal 713104, India}

\begin{abstract}
 We study the possibility of the formation of primordial black holes (PBHs) using an inflaton potential constructed by stabilizing all K\"{a}hler moduli employing both the perturbative and non-perturbative corrections in the large volume scenario in type-IIB/F string theory. We further introduce a small perturbation in one of the K\"{a}hler moduli that modifies the existing potential in the field space. The effective correction in the moduli space is made by introducing a localised Gaussian bump,  which induces an ultra-slow-roll feature in the inflaton potential during the early cosmological inflation. The evolution of the amplitude of the curvature power spectrum is solved numerically by using the Mukhanov-Sasaki equation, with Bunch-Davies initial conditions. The inclusion of the Gaussian bump produces a sustainable enhancement in the amplitude of curvature power spectra around $k\sim 1.12 \times 10^{9} Mpc^{-1}$ for the formation of PBHs. The abundance of PBHs ($f_{PBH}$) is calculated using the Press-Schechter formalism for different values of critical density contrast and presented in the background of contemporary $f_{PBH}$ bounds and checked for viability with existing microlensing bounds, corresponding to a mass value $10^{-6}M_\odot$ for $\delta$ threshold value $\delta_c=0.439$. The Hawking temperature and evaporation time are also determined for these black holes and found to be of the order of  $2.78\times 10^{-15}$ GeV and $8.5\times10^{38}$ Gyr, respectively.  The enhancement of curvature power spectra lies in the sensitivity region of the SKA observations, suggesting an indication that scalar-induced gravity waves may be probed in future observations.
\end{abstract}

\maketitle

%\tableofcontents

\section{\label{sec:level1} Introduction}
At present, string theory is believed to be the most fundamental theory to explain the origin of the universe \cite{Chernoff:2014cba}. There is, however, some tension between cosmological inflation and string theory, as it is difficult to achieve a de-sitter (dS) minima from the latter. String theory operates in the AdS vacuum, whereas a positive cosmological constant is required for inflation, i.e., a dS vacuum \cite{Linde:2017pwt} is needed. In the KKLT formalism \cite{Kachru:2003aw}, an anti-D3 brane lifts the AdS minimum to a metastable dS minimum, after non-perturbative quantum correction to the superpotential. Following stringy cosmology, an inflaton potential can be constructed by stabilizing all the K\"{a}hler moduli through perturbative and non-perturbative corrections in the large volume scenario with a geometry of three intersecting magnetized D7 branes compactified on a 6D $T^6/Z_n$ orbifold of a Calabi-Yau threefold. In ref. \cite{Let:2023dtb, sarkar2025study}, a general scenario has been considered in which two non-perturbative contributions are taken into account. The non-perturbative corrections come from either Gaugino condensation on stacks of D7 branes or the instanton effect, which includes K\"ahler moduli in the flux-induced superpotential \cite{Kachru:2003aw}. The perturbative corrections arise from multi-graviton scattering amplitudes taken up to one-loop level. The logarithmic correction comes from the exchange of $KK$ excitations from localized gravity sources on the loop to the  D7 branes/orientifold planes in the large volume limit \cite{antoniadis2020logarithmic}. The perturbative corrections modify the K\"{a}hler potential, which can break the no-scale structure. In this framework, the D-term contribution, which comes from the magnetized D7 brane configuration, uplifts the $AdS$ vacua to the $dS$ ones. After achieving the dS vacua, the resulting scalar potential can be utilized to study the dynamics of early cosmic inflation.  \par In our recent study \cite{Let:2023dtb, sarkar2025study} of early inflation, we have used a stringy inflaton potential  derived from IIB/F theory  given by
\begin{equation}
V(\phi)= V_0( p e^{\frac{-3\phi}{\sqrt{2}}}[\phi-q e^{\frac{\phi}{\sqrt{2}}}-s\phi e^{\frac{-\phi}{\sqrt{2}}} + x]+ \lambda).
\label{eq:V-inf1}
\end{equation}
The potential parameters are given in the table as
\begin{table}[h]
    \centering
     \caption{Parameter values of the stringy inflaton potential that yield a successful inflationary scenario taken from \cite{ sarkar2025study} }
    \begin{tabular}{|c|c|c|c|c|c|}
    \hline
    $V_{0}$ &p & q & s & x & $\lambda$  \\\hline
   $5\times10^{-5}$ & $-0.690$ & $-2.020 $ & $182.14$ &$-441.241$ & $4.416\times10^{-6}$ \\\hline
    \end{tabular}
   \label{tab:table1}
\end{table} 
Here, $V_0=5\times 10^{-5}$ is taken to make the potential consistent with the CMB data and follows the COBE normalization. In our previous study \cite{sarkar2025study}, the normalization was obtained using the slow-roll approximation in time space, for which $(V_0=10^{-4})$ was adopted. 
The important feature of this potential is that it provides slow-roll inflation and supports the prescribed value of the scalar power spectrum ($P_{\mathcal{R}} (k)$$\sim10^{-9}$), tensor-to-scalar ratio ($r=0.0011$), and scalar spectral index ($n_s=0.9647$) reported in the Planck 2018 data \cite{Planck:2018jri, Planck:2018vyg}. The successful realization of the slow-rolling phenomenon and cosmological inflation using the stringy potential (\ref{eq:V-inf1}) \cite{sarkar2025study}  may raise an important question: Can the same inflaton potential provide a regime of the ultra-slow-roll (USR)  phase and extend the applicability of the string theoretic inflaton potentials \cite{ Let:2023dtb, sarkar2025study} to the arena of formation of PBHs?  
\par The formation of PBHs is believed to be associated with the collapse of large density fluctuations in the early universe \cite{1967SvA....10..602Z, carr1974black, Musco:2020jjb, musco2019threshold}. Inflation incorporates the most compelling mechanism for generating the primordial curvature perturbations that seed the large-scale structure of the Universe. Zel'dovich and Novikov \cite{1967SvA....10..602Z} first proposed that the formation of PBHs originates from primordial density fluctuations and was later developed by Hawking and Carr \cite{carr1974black}. Following the arrival of inflationary theory, it was recognized in the early 1980s that quantum fluctuations generated during inflation could be amplified and subsequently collapse upon horizon re-entry, leading to the formation of primordial black holes. It is generally believed that deviation from the slow-roll behavior is necessary for the production of PBHs. Formation of PBHs is usually signaled by an enhancement in the curvature power spectrum, $P_{\mathcal{R}} (k)$.  The slow-roll potentials have a  $P_{\mathcal{R}} (k)$ $\sim$ $10^{-9}$  at the  Cosmic Microwave Background (CMB)  scale \cite{Planck:2018vyg,Planck:2018jri}. The enhancement to the order of $10^{-2}$ in $P_{\mathcal{R}} (k)$  is considered as the signal of PBH formation \cite{Young:2019osy}.  PBH production has recently been studied in the USR framework for superconformal inflationary $\alpha$-attractor \cite{Mahbub:2019uhl}, improved E-model $\alpha$-attractor \cite{Frolovsky:2023hqd}, modulated chaotic $\alpha$-attractors \cite{dalianis2019primordial} and string-theoretic models \cite{ozsoy2018mechanisms}. The amplification in the curvature power spectrum can also be achieved by the Higgs inflation model \cite{Ezquiaga:2017fvi}, hybrid inflation \cite{Garcia-Bellido:1996mdl}, warm inflation \cite{arya2020formation}, and the quantum diffusion model \cite{biagetti2018primordial}. In this context, PBHs have been a subject of theoretical interest for the last five decades or so \cite{carr1974black, PhysRevD.40.1725, Musco:2020jjb, Young:2019osy, Mahbub:2019uhl, Frolovsky:2023hqd, dalianis2019primordial, Ezquiaga:2017fvi, ozsoy2018mechanisms, Garcia-Bellido:1996mdl, arya2020formation, biagetti2018primordial, Arbey:2024ujg,carr2020primordial, carr2018primordial, cappelluti2022exploring, laha2019primordial, green2021primordial}. Recently,  considerable attention has been drawn to various ongoing and proposed experimental projects (for a recent review, see Ref. \cite{bagui2025primordial}) concerning these interesting light/ultra-light and massive objects.  Of late, light \cite{Franciolini:2022htd, calza2024determining} and ultralight \cite{datta2021baryogenesis, Papanikolaou:2023oxq, Bhaumik:2023wmw, Ray:2022rfz, profumo2025ultralight, Guo:2024odw} PBHs and their various properties are also getting particular attention. Light and ultra-light PBHs shed light on events such as Baryogenesis \cite{datta2021baryogenesis} and $H_{0}$ tensions \cite{Papanikolaou:2023oxq}. Light PBHs have a short lifetime compared to the age of the universe and evaporate through Hawking radiation.  On the contrary, the massive PBHs do not evaporate by Hawking radiation yet and have a lifetime longer than the age of the present universe and provide a probe to the early universe.  Among various possible mass windows, those PBHs that have not evaporated yet have gained considerable attention as a plausible dark matter (DM) candidate \cite{carr1974black, PhysRevD.40.1725, Musco:2020jjb, Young:2019osy, Mahbub:2019uhl, Frolovsky:2023hqd, dalianis2019primordial, Ezquiaga:2017fvi, ozsoy2018mechanisms, Garcia-Bellido:1996mdl, arya2020formation, biagetti2018primordial, Arbey:2024ujg, carr2020primordial}. 
\par So far in the above text, we have discussed the enhancement of the amplitude of the curvature power spectrum and ultra-slow rolling of the inflaton field that causes the dynamical collapse for the formation of PBH. The most common practice has been to modify the slow-roll inflaton potential itself in order to generate a region of either USR \cite{Mahbub:2019uhl, ozsoy2018mechanisms, choudhury2024no, PhysRevD.94.103515, dimopoulos2017ultra, ragavendra2021primordial, garcia2017primordial} or a point of inflection \cite{Frolovsky:2023hqd, Choudhury:2013woa, ballesteros2018primordial, bhaumik2020primordial}, which would enhance (spike) the curvature power spectrum, signaling the formation of PBH. In the above literature survey, we have seen that the small USR region in the inflaton potential was achieved by modifying the shape of the potential through introducing a bump or dip \cite{Mishra:2019pzq, Yang:2024ntt}, a step feature, heavy field back reaction, or waterfall field \cite{Tada:2023fvd, FUMAGALLI2023137921}. 
 After a careful and thorough literature review \cite{Mahbub:2019uhl, ozsoy2018mechanisms, choudhury2024no, PhysRevD.94.103515, dimopoulos2017ultra, ragavendra2021primordial, garcia2017primordial, Frolovsky:2023hqd, Choudhury:2013woa, ballesteros2018primordial, bhaumik2020primordial, Tada:2023fvd, FUMAGALLI2023137921, Mishra:2019pzq, Yang:2024ntt}, we note that the origin of the small USR region from an underlying fundamental theory has hardly been discussed.  In the USR region, these corrections can be approximated by a Gaussian bump, thereby introducing the USR transition of the inflaton field during inflation. This provides a physically motivated origin of the USR instead of the phenomenological modification.  
In this context, it is to be noted that, in ref. \cite{Mishra:2019pzq}, the Gaussian bump of the form $ A exp[\frac{-1}{2}\frac{(\phi-\phi_{0})^2}{\delta^2}]$ is incorporated in the KKLT-type potential, which originates from the non-perturbative Kähler moduli stabilization mechanism \cite{kachru2003towards}, to introduce a USR phase to study the formation of PBH.  However, in this study \cite{Mishra:2019pzq}, the KKLT-type potential employs a top-down approach; the mathematical origin of the aforementioned localized features is not explicitly addressed. To achieve this enhancement in the amplitude of the curvature power spectrum there are many successful phenomenological models that generates the phase of USR \cite{Mahbub:2019uhl, ozsoy2018mechanisms, choudhury2024no, PhysRevD.94.103515, dimopoulos2017ultra, ragavendra2021primordial, garcia2017primordial, Mishra:2019pzq, Yang:2024ntt} or point of inflection \cite{Frolovsky:2023hqd, Choudhury:2013woa, ballesteros2018primordial, bhaumik2020primordial}. However, the microscopic origin of these localized features from an underlying theory is largely unexplored and has hardly been discussed in the literature \cite{Mahbub:2019uhl, ozsoy2018mechanisms, choudhury2024no, PhysRevD.94.103515, dimopoulos2017ultra, ragavendra2021primordial, garcia2017primordial, Mishra:2019pzq, Yang:2024ntt, Frolovsky:2023hqd, Choudhury:2013woa, ballesteros2018primordial, bhaumik2020primordial}. We make use of a rigorously derived string-theoretic slow-roll inflaton potential $V(\phi)$ given in (\ref{eq:V-inf1}) and a plausible source of the origin of a Gaussian bump, which induces a USR phase in the potential. In contrast to the KKLT-type potential \cite{kachru2003towards, Mishra:2019pzq}, the potential in Eq. (\ref{eq:V-inf1}) is stabilized through both perturbative and non-perturbative corrections in the Large Volume scenario (LVS). The introduction of the Gaussian bump in this potential will explore a different kind of string geometry and provide an additional motivation for the generation of a temporary USR phase. In addition, it is well known that the early universe and DM are deeply connected with PBHs \cite{Green:2020jor, carr2022primordial, green2024primordial}, as they may arise from sufficiently large density perturbations during the post-inflationary evolution. In particular, the enhancement of the curvature perturbations due to the introduction of the USR phase, on small scales, can produce overdense regions, which, upon re-entering the horizon after inflation, can produce PBHs. These PBHs are cold, non-baryonic, and, if formed in the right abundance, they may behave as a good plausible candidate for the DM \cite{green2024primordial}. Another advantage is that, like axions or WIMPs, PBHs are not a new particle \cite{green2024primordial}. Moreover, unlike particle dark-matter candidates, the mechanism of PBH formation is the collapse of sufficiently large primordial density fluctuations in the early Universe, which provides a direct connection between early inflationary dynamics and the DM sector. In this context, it is noteworthy that the mass of the PBH, namely $M_{PBH}$, and their abundance, namely $f_{PBH}$, are two definitive parameters that define and constrain PBHs formed during the early inflationary epoch.  The mass of the PBH is formed when the enhanced perturbations re-enter the horizon, which directly links the early inflationary dynamics with the mass spectrum of the PBH. The abundance of the PBHs, namely the $f_{\rm PBH}$, measures the contribution of the produced PBHs to the present-day dark matter density. Therefore, it is essential to investigate the formation of the PBH during the USR phase, using the inflaton potential presented in Eq. (\ref{eq:V-inf1}) with particular emphasis on the theoretical origin of the localized feature. Additionally, both quantities $M_{PBH}$ and  $f_{PBH}$ are investigated to check whether our stringy model, in which the USR phase is introduced by deforming one of the K\"ahler moduli, can lead to a relevant amount of PBH production.
 \par  The paper is organized as follows: In Sec. \ref{sec:level2},
we present the origin of the localized bump from wrapped geometry in type IIB/F theory. In Sec. \ref{sec:level3}, we examine the dynamics of the early inflation and show how the bump, i.e., a localized Gaussian feature, brings a transient USR phase. We then show the behavior of scalar perturbations by solving the Mukhanov–Sasaki equation and compute the corresponding primordial power spectrum. In addition, we also present the behaviour of the slow-roll parameters, highlighting the enhancement of the scalar power spectrum and implications for primordial black hole formation. In Sec. \ref{sec:fpbh},  the formation and present abundance of primordial black holes $f_{PBH}$, including their characteristic mass $M_{\rm PBH}$ along with their evaporation time and Hawking temperature, are discussed. Finally, Sec. V summarizes our main results and presents our concluding remarks. 
\section{\label{sec:level2} Origin of the Gaussian bump}
 From a top-down theoretical perspective, the realization of cosmic inflation, characterized by the rapid expansion of the Universe, can be achieved through various mechanisms depending on the underlying geometrical origin of the inflaton field and the resulting scalar potential \cite{Baumann:2014nda}.
 The most prevalent mechanisms that could be responsible include: (i) Separation between brane-anti-brane pairs, which are attracted by the gravitational interactions and  R-R four-form potential C4, may serve as a viable candidate for the inflaton \cite{Dvali:1998pa, kachru2003towards}. The mobile branes, generally the D3 branes, move radially inwards through the throat towards their anti-brane counterpart or D7 brane. The separation between them acts as the inflaton, and the attractive interaction generates the effective inflaton potential. When the open string connecting the brane-anti-brane pair becomes tachyonic and condensed, transferring the energy to the reheating process, the inflation ends,  (ii) an axionic field originating from shift symmetry valid up to all orders in perturbative theory, along with weakly broken symmetry by non-perturbative effects or the presence of a brane, leads to axion inflation \cite{Baumann:2014nda, Flauger:2009ab}, (iii) one of the Kähler moduli which determine the volume of the internal four-cycles can act as inflaton. Each of these mechanisms arising from the different internal geometries is capable of producing a different kind of inflationary mechanism and a distinct four-dimensional effective theory of early cosmic inflation. Unlike brane or axionic inflation, the potential in Eq. (\ref{eq:V-inf1}) combines perturbative and non-perturbative corrections in the LVS to stabilize all the k\"ahler moduli with a geometry of three intersecting D7-brane compactified on a 6d $T^6/Z_n$ orbifold of Calabi-Yau threefold $(CY_3)$ \cite{Green:1997as}.
 The compactification on the $T^6/Z_{n}$ orbifold subsequently raises some scalar fields known as k\"ahler moduli. In the present construction of the potential in Eq. (\ref{eq:V-inf1}), the internal volume of the compactified manifold can be written as \cite{Basiouris:2021sdf, Basiouris:2020jgp}
 \begin{equation}
 \mathcal{V}=\sqrt{\tau_1\tau_2\tau_3}
 \end{equation}
 Among the three Kähler moduli $\tau_{1}$, $\tau_{2}$ and $\tau_{3}$, two moduli $\tau_{1}$ and $\tau_{2}$ are constant, i.e., the covariant derivative of the superpotential is zero with regard to these two moduli. The modulus $\tau_3$, after stabilizing by perturbative corrections, provides the inflaton field and remains dynamical. This is manifested as a logarithm of $CY_3$ volume =$\sqrt{2} ln \mathcal{V} =\frac{1}{\sqrt{2} } ln(\tau_{1}\tau_{2}\tau_{3})$ and identified with the inflaton field  $\phi$  after canonical normalization.  In this context, it is noteworthy that the geometrical configuration in the construction of the potential in Eq. (\ref{eq:V-inf1}), specifically the $D$-term, a logarithmic correction arising from the stacks of three intersecting D7-branes, lifts the vacuum from AdS to dS space, ensuring a positive cosmological constant, a necessary condition for early cosmic inflation \cite{sarkar2025study}. The effective four-dimensional potential in (\ref{eq:V-inf1}) receives contributions from both the D-term potential and F-term potential. The F-term potential is determined by the Kähler potential and the superpotential through the standard $(\mathcal{N}=1)$ supergravity F-term expression \cite{Basiouris:2021sdf, Basiouris:2020jgp},
\begin{equation}
 V_F=e^K\left(K^{I\bar J}D_IWD_{\bar J}\bar W-3|W|^2\right), 
 \label{eq:F-term}
\end{equation}
where the quantum-corrected K\"ahler potential is given by \cite{Basiouris:2020jgp, Basiouris:2021sdf, Let:2023dtb}
\begin{equation}
K^0=-2\ln(\mathcal{V}+\delta)    
\end{equation}
with
\begin{equation}
\delta = \xi + \sum_{j=1}^{3} \eta_j \ln(\tau_j)
= \xi + \eta \ln(\tau_1 \tau_2 \tau_3).    
\label{eq:delta}
\end{equation}
 The logarithmic correction depends explicitly on the Kähler moduli, which establishes a direct connection between the internal geometry of the compactified manifold and the four-dimensional scalar potential. 
 \par In the present work, we would like to investigate a plausible way of generating a region of USR by incorporating a localized correction in the existing potential (\ref{eq:V-inf1}) from the geometry of the internal manifolds, and the correction depends on the internal manifold and corrected K\"ahler potential. To do such a correction in the existing potential, we perturb one of the K\"ahler moduli, i.e., specifically $\tau_3$, which produces the inflaton field $\phi$, as follows
\begin{equation}
    \tau_{(3)}= \tau_{3}^{0}+\delta\tau_{3}^{0} \quad\left|\delta\tau_{3}^{0}\right|<<\tau_{3}^{0}.
\end{equation}
The perturbations to the K\"ahler moduli $\tau_3$ are so small that they do not destabilize the compactification, i.e., heavy moduli remain fixed at their stabilized values. The perturbation represents a small deformation around the stabilized vacuum configuration which may arise from various higher-order corrections \cite{Baumann:2009ds}. Any perturbation to the stabilized K\"{a}hler moduli propagates into the effective inflaton potential through the Kähler potential. By substituting the perturbed modulus $\tau_3$  into Eq.         (\ref{eq:delta}), i.e., the logarithmic correction becomes   
\begin{equation}
    \delta\rightarrow\delta'
= \xi + \eta \ln[\tau_1^{0} \tau_2^{0} (\tau_3^{0}+\delta\tau_{3})], 
\end{equation}
which induces a corresponding correction to the K\"{a}hler potential
\begin{equation}
    K=K^{0}+\Delta 
    K
\end{equation}
Since the exponential factor $e^k$ in Eq. (\ref{eq:F-term}), with the inverse K\"{a}hler metric of  the K\"{a}hler potential determining the scalar potential, the perturbations in the K\"{a}hler moduli explicitly propagate in the scalar potential through the F-term potential, which introduces a correction term to the original slow-roll potential written as 
\begin{equation}
    V_m(\phi)= V_{SR}(\phi)+\Delta V(\phi)
    \label{eq:totalpotential}
\end{equation}
where the subscript `SR' in (\ref{eq:totalpotential}) denotes the slow-roll potential given in (\ref{eq:V-inf1}) and $\Delta V(\phi)$ represents the localized correction induced by the perturbation applied on the $\tau_3$ moduli \cite{Baumann:2009ds}. 
\par We have just seen that the introduction of the perturbation in one of the K\"ahler moduli leads to the existence of the correction $\Delta V(\phi)$ in the potential. However, the exact mathematical form of the localized correction cannot be determined without a comprehensive treatment of the higher-order correction to compactification. To introduce a localized correction, we employ an effective toy model description.  The perturbation in the $\tau_3$ moduli modifies the k\"ahler potential itself; thus the localized correction $\Delta V$ is expected to be a deformation of the same energy scale as the existing slow-roll potential instead of an independent one. We, therefore, parametrize the correction as 
\begin{equation}
    \Delta V(\phi)= V_{SR}(\phi)\epsilon(\phi),
\end{equation}
where the simplest smooth localized profile 
\begin{equation}
    \epsilon(\phi)=A exp[-\frac{1}{2} \frac{(\phi-\phi_{0})^2}{\Delta^2}],
    \label{eq:epsilon}
\end{equation}
introduces a localized correction in the slow-roll potential. Here, $A$, $\phi_{0}$, and $\delta$ denote the amplitude, position, and width of the localized Gaussian bump.
\par  The use of the Gaussian bump (\ref{eq:epsilon}) in Eq. (\ref{eq:totalpotential}) reshapes the slow-roll potential and takes the following form 
 \begin{equation}
 \begin{split}
V_m(\phi)=& V_0( p e^{\frac{-3\phi}{\sqrt{2}}}[\phi-q e^{\frac{\phi}{\sqrt{2}}}-s\phi e^{\frac{-\phi}{\sqrt{2}}} + x]+ \lambda) \times \\& (1+ A exp[\frac{-1}{2}\frac{(\phi-\phi_{0})^2}{\Delta^2}]) .
\end{split}
\label{Vmodified}
\end{equation} 
\begin{figure}[h]
    \centering
    \includegraphics[width=0.5\textwidth]{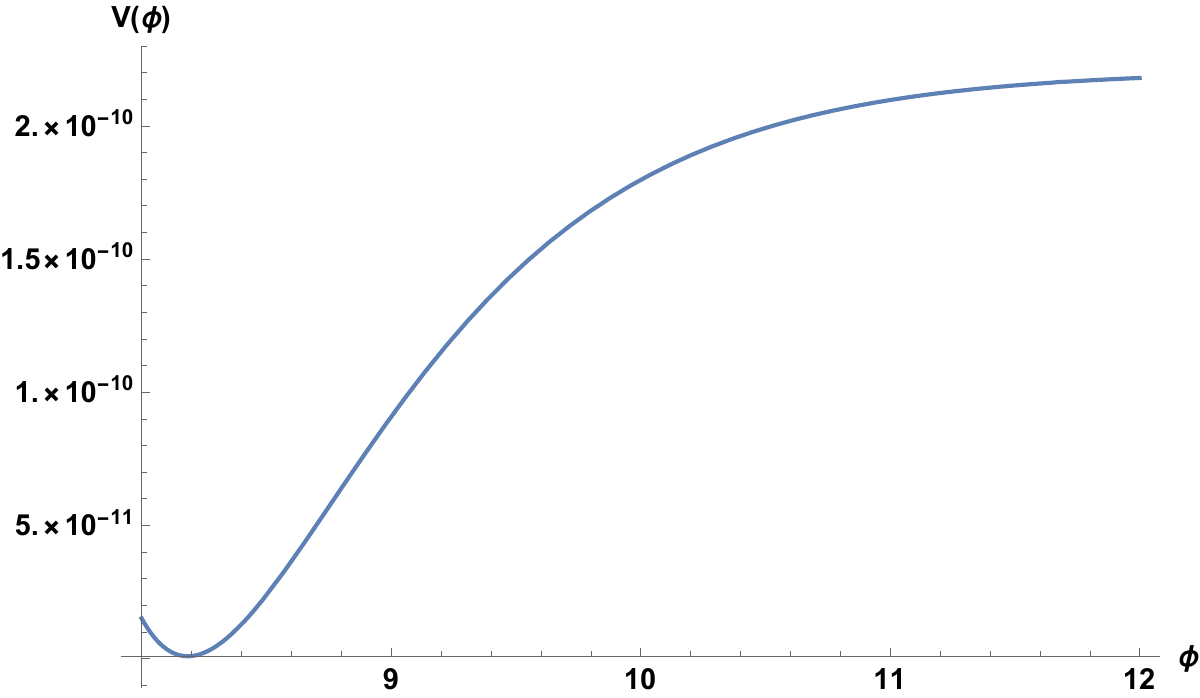}
    \caption{Plot of the modified potential from Eq. (\ref{Vmodified}) v/s the canonically normalized field $\phi$. using the parameters described in Table \ref{tab:table1} and \ref{tab:table2}}
    \label{fig:originalpotential}
    \end{figure}
    \begin{figure}[h]
    \centering
    \includegraphics[width=0.5\textwidth]{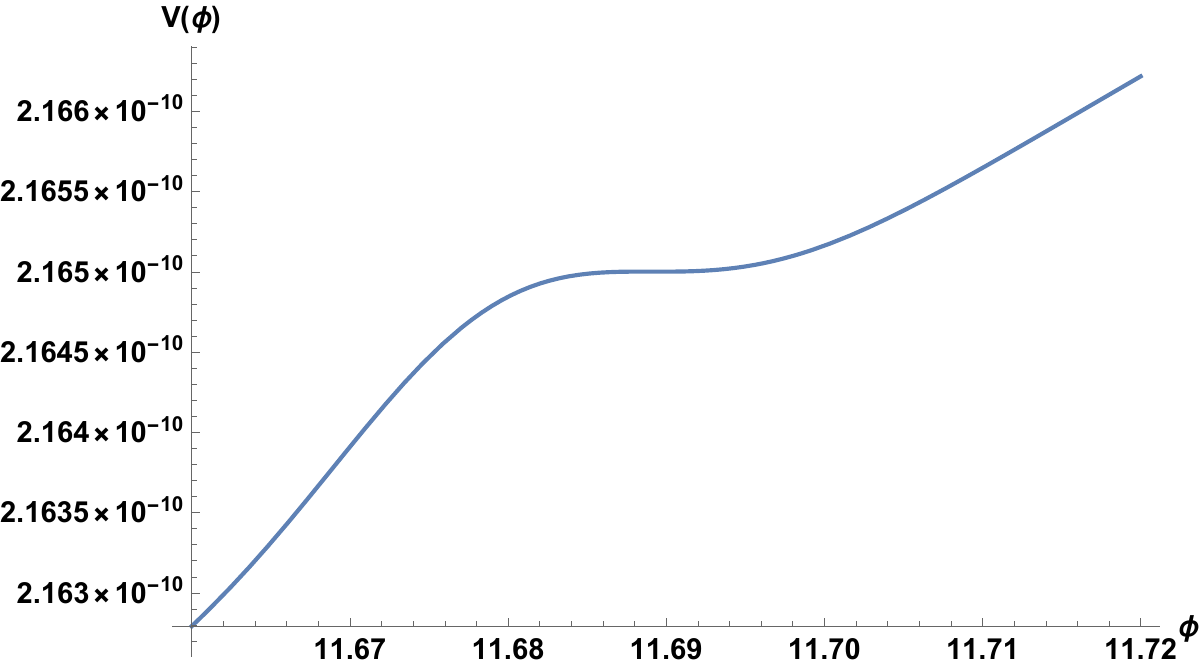}
    \caption{A magnified view of the localized region of the modified inflaton potential shown in Fig. \ref{fig:originalpotential} depicts the Gaussian bump that is induced in the original potential and leads the transient to the ultra-slow-roll phase}
    \label{fig:bumpofpotential}
    \end{figure}
    \begin{figure}
        \centering
        \includegraphics[width=0.5\textwidth]{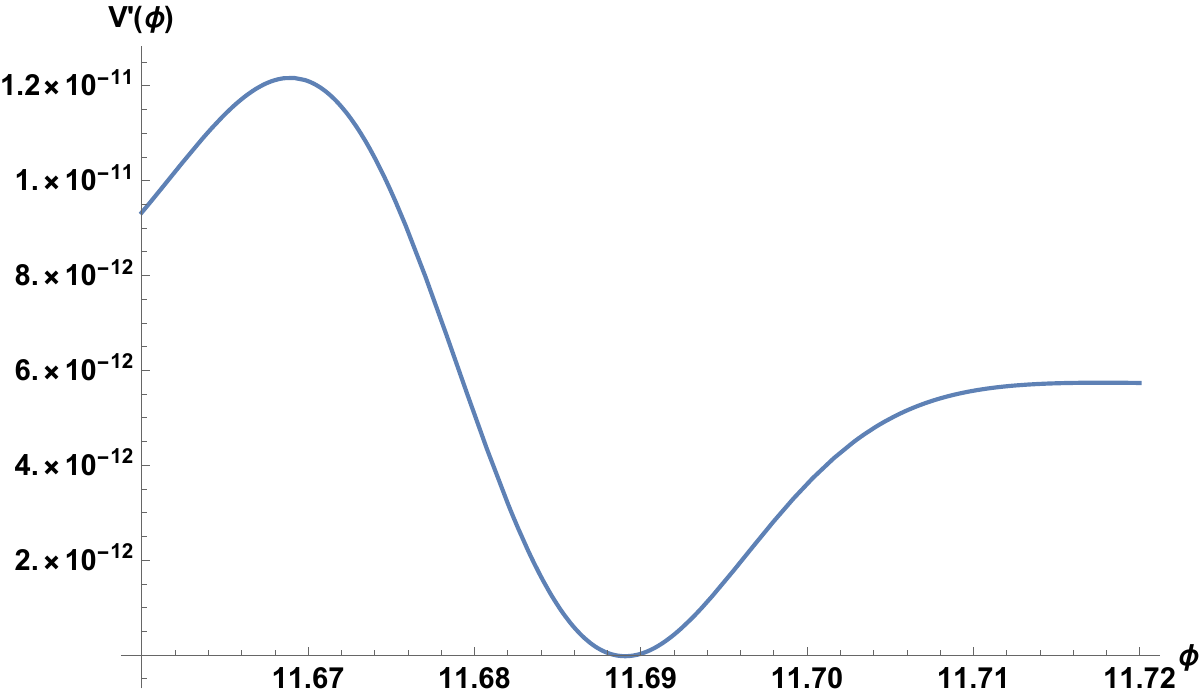}
        \caption{Variation of the first derivative of the modified inflaton potential in Eq. (\ref{Vmodified}) v/s scalar field $\phi$ using the parameters described in Table \ref{tab:table1} and \ref{tab:table2}.   }
        \label{fig:potential derivative}
    \end{figure}
\begin{table}[ht]
\centering
\caption{Parameters characterizing the localized Gaussian correction used in the numerical analysis}
\label{tab:gaussian_parameters}
\begin{tabular}{c c}
\hline
Parameter & Value \\
\hline
$A$ & $4.5815
\times10^{-4}$ \\
$\phi_{0}$ & $11.6790$ \\
$\Delta$ & $1.00\times10^{-2}$ \\
\hline
\label{tab:table2}
\end{tabular}
\end{table}
Using the parameter sets given in Tables \ref{tab:table1} and \ref{tab:table2}, we numerically evaluate the resulting inflaton potential, as shown in Figs. \ref{fig:originalpotential} and \ref{fig:bumpofpotential}. In this context, we emphasize that, unlike the parameters entering the underlying string-motivated slow-roll potential, the parameters A, $\phi_0$, and $\Delta$ listed in Table \ref{tab:table2} are not derived from a microscopic string-theoretic calculation. These parameters are phenomenological parameters introduced within the effective toy-model description of the localized correction. Figure \ref{fig:originalpotential} displays the overall form of the modified potential, while Fig. \ref{fig:bumpofpotential} provides a magnified view of the localized region containing the Gaussian feature. Figure \ref{fig:potential derivative} shows that the Gaussian feature also modifies the slope of the stringy potential around $\phi \simeq \phi_{0}=11.6790$ (Table \ref{tab:table2}). When the inflaton rolls down near this region of Fig. \ref{fig:bumpofpotential}, the magnitude of the slope of the potential is subsequently suppressed, resulting in a temporary slower rolling of the inflaton around the region, so-called the ultra-slow-rolling of the inflaton field. It again joins hands with the early slow-rolling process at $\phi\simeq11.6688$, following the SR-USR-SR process. The parameter values presented in Table \ref{tab:table2} are so chosen to generate a localized feature capable of producing the required enhancement of the amplitude of the primordial curvature power spectrum.  The resulting enhancement and its implications for primordial black hole production are investigated quantitatively in the subsequent sections.
\section{ \label{sec:level3}Ultra-slow-roll dynamics and curvature perturbations}During the period of inflation, the comoving Hubble radius, represented as $(aH)^{-1}$, decreases as the scale factor accelerates. Thus any quantum perturbation characterized by a wavenumber $k$ begins subhorizon-sized $(k>aH)$ and subsequently crosses outside the horizon ($k<aH$, super-horizon) \cite{Kolb:1990vq}. The perturbation then freezes. At the end of the early inflationary epoch, when the Hubble horizon begins expanding again, the modes start to re-enter the horizon. In the context of PBH formation, a transition from the SR-USR-SR phase during inflation causes enhancement of the amplitude of curvature power spectra from $10^{-9}$ (SR) to $10^{-2}$ at the USR phase, providing the mode of large-amplitude primordial fluctuations \cite{franciolini2022primordial, Mahbub:2019uhl, merchand2026exploring}. These large fluctuations of curvature power spectra ultimately act as the seeds for the formation of the PBHs \cite{1967SvA....10..602Z, carr1974black}.
The formation of PBHs is commonly studied in momentum space by observing an enhancement of the amplitude of the curvature power spectrum ${P}_\mathcal{R}(k)$. ${P}_\mathcal{R}(k)$ can be obtained as a solution of the Mukhanov-Sasaki (MS) equation \cite{Musco:2020jjb,  Mahbub:2019uhl, Mukhanov:1988jd, Leach:2001zf} for a particular inflaton potential with the Bunch-Davies (BD)  initial conditions \cite{Bunch:1978yq}. In our recent paper \cite{sarkar2025study}, we have found the amplitude of the scalar power spectrum and studied early slow-roll inflation using the potential in Eq. (\ref{eq:V-inf1}). In the present work, we see the effects on slow-roll dynamics and the curvature perturbation after the introduction of a Gaussian bump in the inflaton potential using (\ref{Vmodified}). 
\par Let us first briefly introduce the curvature power spectrum $P_{\mathcal{R}}(k)$ and then the applicability of the MS equation in solving $P_{\mathcal{R}}(k)$ in the BD vacua. The Einstein-Hilbert (EH) action, having a minimal coupling between the inflaton field ($\phi$) and the background metric ($g^{\alpha\beta}$), can be written as,   
\begin{equation}
    S=\int d^4 x \sqrt{-g}\left(\frac{1}{2}R -\frac{1}{2}g^{\alpha\beta}\partial_{\alpha}\phi\partial_{\beta}\phi -V(\phi)\right),
\end{equation}
where $V(\phi)$ is the inflaton potential. We shall be working in Planck units where $M_{pl}=1$.
If the action is minimized with respect to the inflaton field $\phi$, we get an equation of motion, also known as the cosmological  Klein-Gordon equation, which is given by \cite{Baumann:2009ds} 
\begin{equation}
        \ddot{{\phi}} +3H \dot{{\phi}}+ \frac{\partial{V}}{\partial{{\phi}}}=0
        \label{eq:kleingordan}
    \end{equation} 
    where H is the Hubble parameter. To determine various parameters that are necessary to describe inflation, we convert Eq. (\ref{eq:kleingordan}) in terms of e-folds using $dN_{e}=Hdt$ as \cite{Mahbub:2019uhl}
    \begin{equation}
        \frac{d^2{\phi}}{dN_{e}^2}+ 3\frac{d{\phi}}{dN_{e}}-\frac{1}{2}\left(\frac{d{\phi}}{dN_{e}}\right)^3+\left[3-\frac{1}{2}\left(\frac{d{\phi}}{dN_{e}}\right)^2\right]\frac{\partial ln V}{\partial{\phi}}=0.
        \label{eq:kginn}
    \end{equation}
    Inflation ends when the first slow-roll inflationary parameter becomes \cite{Mahbub:2019uhl},
    \begin{equation}
     {\epsilon_{i}}=\frac{1}{2}\left(\frac{d{\phi}}{dN_{e}}\right)^2=1. 
     \label{eq:firstslowroll}
    \end{equation}
    The second slow-roll inflationary parameter is defined as,
    \begin{equation}
        {\eta_{i}}=-\frac{\ddot{\phi}}{H\dot{\phi}}={\epsilon_{i}}-\frac{1}{2}\frac{d ln {\epsilon_{i}}}{dN_{e}}.
        \label{eq:secondslowroll}
    \end{equation}
    During slow-roll inflation, $\epsilon_{i}$ and $\eta_{i}$ remain very less than the unity. The small value of the second slow-roll parameter ensures that the change in  $\epsilon_{i}$  with respect to e-folds is small.
    The curvature power spectrum can be  defined  under the slow-roll approximation as \cite{Baumann:2009ds, Mahbub:2019uhl},
     \begin{equation}
       {P}_{\mathcal{R}}(k)= \frac{H^2}{8\pi^2\epsilon_{i}}.
    \label{equ:curvaturepowespectrum}
        \end{equation}

During slow-roll inflation, the Hubble parameter remains constant and acts as a resistance term. If we calculate the curvature power spectrum using Eq. (\ref{equ:curvaturepowespectrum}), we will not get an accurate result as the Hubble parameter is not constant during the USR phase. In this phase, the cosmological Klein-Gordon equation becomes $\ddot{{\phi}} +3H \dot{{\phi}} \approx 0 $
    , and the second slow-roll parameter becomes $\eta_{i} = -\frac{\ddot{\phi}}{H\dot{\phi}} \approx 3$.
The Mukhanov-Sasaki equation shows the evolution of the curvature perturbation in $k$-space given by \cite{biagetti2018primordial}
\begin{equation}
    \mathcal{R}''_{k}+2\frac{z'}{z}\mathcal{R}'_{k}+k^2\mathcal{R}_{k}=0
    \label{eq:originalmukhanov}
\end{equation}
where $z=\frac{a {\phi}'}{{H}}$ and $\frac{z'}{z}=2aH(1+\epsilon_i-\eta_i)$. For the numerical solution of the MS equation, the field is redefined as $v_{k}=z \mathcal{R}_{k}$ and $z=ad{\phi}/dN$, which converts (\ref{eq:originalmukhanov}) into \cite{Mahbub:2019uhl}   
\begin{equation}
     \begin{split}
        &\frac{d^2v_{k}}{dN_{e}^2}+(1-\epsilon_{i})\frac{dv_{k}}{dN_{e}}\\&+\left[\left(\frac{k}{aH}\right)^2+(1+\epsilon_{i}-\eta_{i})(\eta_{i}-2)-\frac{d}{dN_{e}}(\epsilon_{i}-\eta_{i})\right]v_{k}=0.
    \end{split}
    \label{eq:Mukhanov}
\end{equation}
In this context, it should be noted that a change of variable into the conformal time ($\tau$) frame drops the Hubble friction term in (\ref{eq:Mukhanov}); the damped oscillator equation reduces to the harmonic oscillator equation.  In the super-horizon mode $k<<aH$, the solution becomes a classical tachyonic or parametric instability. In contrast, in the sub-horizon mode, i.e., $k>>aH$, the solution of the equation resembles a pure undamped harmonic oscillator, and the perturbations behave as free quantum fields in static Minkowski space. These plane wave solutions $e^{-ik\tau}$, i.e., the actively oscillating modes, help to define the initial vacuum state; i.e., the BD vacuum. The MS equation is solved with the BD vacuum state, which acts as an initial condition given by \cite{Bunch:1978yq}
 \begin{equation}
     v_{k}\rightarrow\frac{e^{-ik\tau}}{\sqrt{2k}},
     \label{eq:bunchdavis}
 \end{equation}   
where $\tau$ denotes the conformal time.  The modes $v_{k}$  initially start evolving from the quantum initial conditions, i.e., the BD vacuum, and ultimately exit the horizon during early cosmic inflaton, i.e., super-horizon-sized, and their evolution freezes out. More precisely, to solve Eq.  (\ref{eq:Mukhanov}), following \cite{Mahbub:2019uhl, franciolini2022primordial}, we use non-zero sub-horizon BD initial conditions
\begin{equation}
    Re(v_{k})=\frac{1}{\sqrt{2k}},  \quad Im(\frac{dv_{k}}{dN})=-\sqrt{\frac{k}{2}} \frac{1}{10^{-2}k}
\end{equation} at $k=100 aH$ and numerically  solve (\ref{eq:Mukhanov}) up to $k=0.01 aH$, following the horizon crossing intermediately to determine the power spectrum defined as 
\begin{equation}
    {P}_\mathcal{R}(k)=\frac{k^3}{2\pi^2}\left|{\frac{v_{k}}{z}}\right|_{k<<aH}^2
     \label{eq:mukhanovpowerspectrum}
      \end{equation}
Here, we make use of Eqs. (\ref{Vmodified}), (\ref{eq:kginn}), (\ref{eq:firstslowroll}), (\ref{eq:secondslowroll}) to determine the solution of the MS equation from (\ref{eq:Mukhanov}), and using that in Eq. (\ref{eq:mukhanovpowerspectrum}), the power spectrum for the curvature perturbations can be determined. It is to be noted that to solve Eq. (\ref{eq:Mukhanov}), we have split the variable $v_{k}$ into the real and the imaginary parts. \par
\begin{figure}[h]
    \centering
    \includegraphics[width=0.5\textwidth]{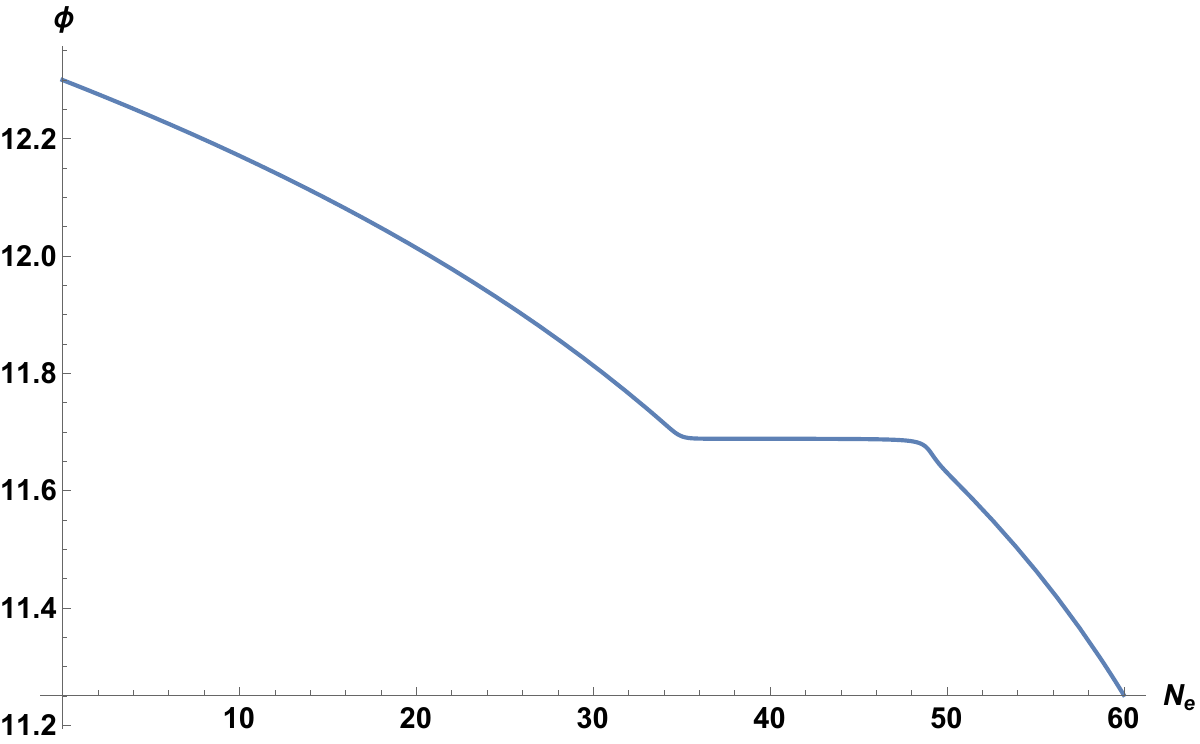}
    \caption{Behaviour of the inflaton field $\phi$ vs the number of e-folds,. Near 35 e-folds the value of $\phi$ becomes constant indicating the USR Phase }
    \label{fig:slow-roll}
\end{figure}

Here, we present a comprehensive summary of the PBH formation for the modified potential presented in Eq. (\ref{Vmodified}). In computing the results for parameters and observables related to PBHs during the USR phase, we use Eqs. (\ref{eq:kginn}) to (\ref{eq:mukhanovpowerspectrum}). 
In Fig. \ref{fig:slow-roll}, we show the behavior of the inflaton field $\phi$  with respect to the e-fold numbers for the modified inflaton potential. It is evident from Fig. \ref{fig:slow-roll} that the USR phase has developed in the potential after 35 e-folds and continues until $\sim 48$ e-folds, and then the slow-roll phase is restored again and $\phi$ moves to its true vacua where the inflation ultimately ends. In this USR phase, the inflaton runs more slowly than the usual slow-roll phase, and the curvature power spectrum starts to increase. After the USR phase ends a\par  

\begin{figure}[h!]
    \centering
\includegraphics[width=0.5\textwidth]{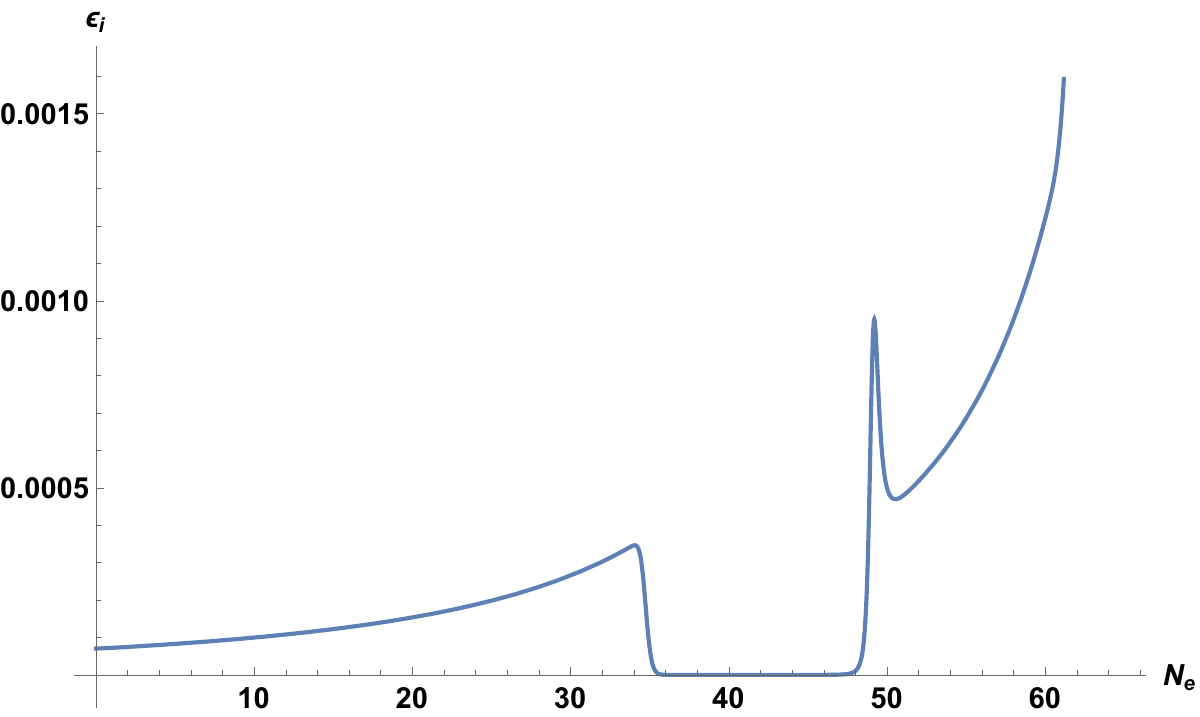}
    \caption{Variation of the first slow-roll parameter vs e-fold from Eq. (\ref{eq:firstslowroll}). The slow-roll parameter breaks down after 35 e-folds, indicating the onset of the ultra-slow-roll phase in the inflation potential.}
    \label{fig:Slow-roll parameter}
\end{figure}

\begin{figure}[h!]
    \centering
\includegraphics[width=0.5\textwidth]{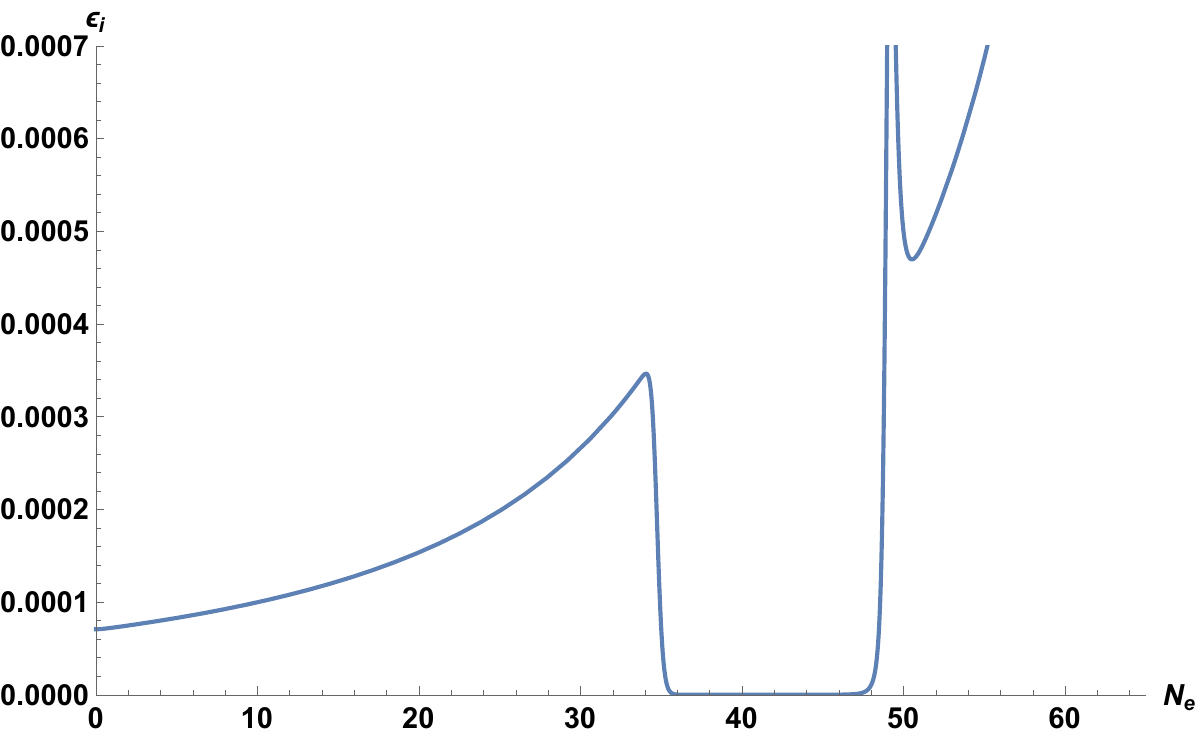}
    \caption{A magnified version of the first slow-roll parameter vs the number of e-folds.}
    \label{fig:Slow-roll parameterd}
\end{figure}

In Fig. \ref{fig:Slow-roll parameter}, we have shown the evolution of the first slow-roll parameter with the number of e-folds for the modified inflaton potential. It is clearly seen from Fig. \ref{fig:Slow-roll parameter} that during the slow-roll phase, the change in the slow-roll parameter ${\epsilon_{i}}$ is very small, and the value increases very slowly. When the inflaton runs on the bump, the slow-roll breaks down, which is one of the essential criteria for PBH formation. We observe a sharp fall in the first slow-roll parameter ${\epsilon_{i}}$, and again it increases after going beyond the bump, and the value ultimately reaches unity in its true vacuum, where the inflation truly ends. We also have shown a magnified plot of the first slow-roll parameter in Fig. \ref{fig:Slow-roll parameterd}.

\begin{figure}[h!]
    \centering
\includegraphics[width=0.5\textwidth]{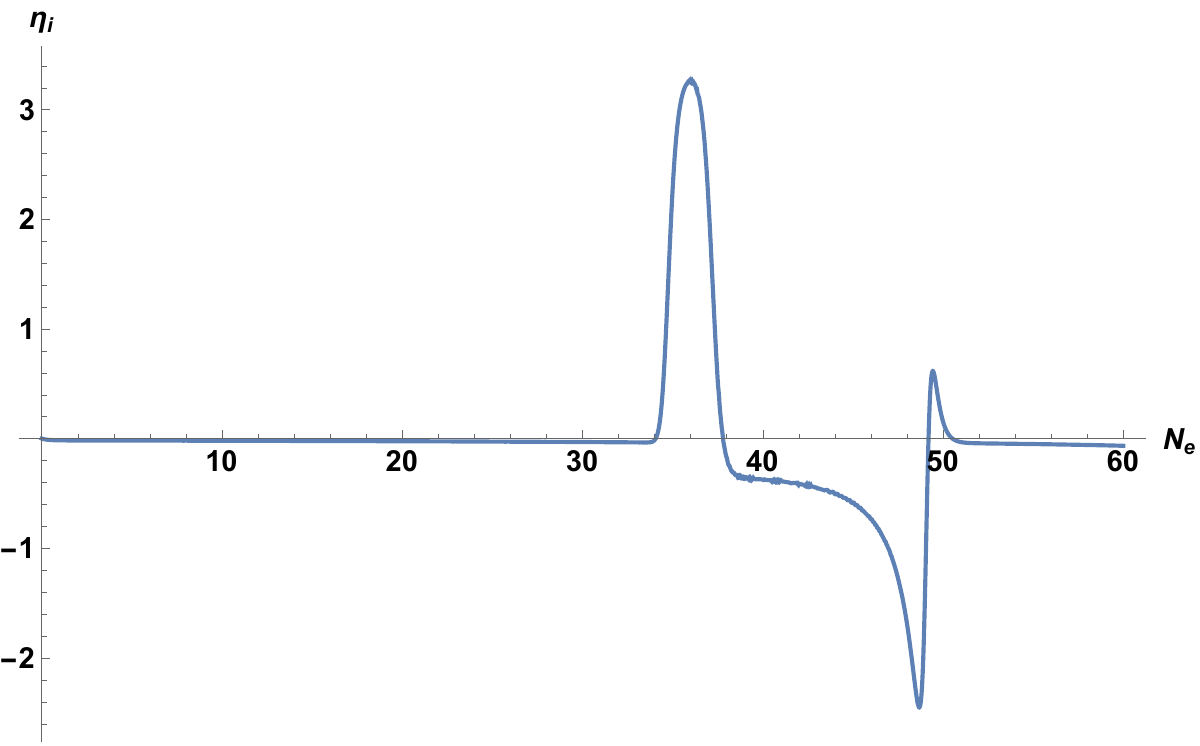}
    \caption{Variation of the second slow-roll parameter vs the number of e-folds. After 35 e-folds, $\eta\simeq 3$ corresponds to the USR phase, during which the inflaton undergoes a strongly suppressed motion}
    \label{fig:sSlow-roll parameterd}
\end{figure}
 In Fig. \ref{fig:sSlow-roll parameterd} we depict the behaviour of the second slow-roll parameter for the modified stringy potential as a function of the number of e-folds. Around the $\sim35$ number of e-folds where the USR phase exists, the second slow-roll parameter reaches a value of $\sim3.2$. This large value supports the USR phase, where the variation of the potential is almost zero,  and Eq. (\ref{eq:kleingordan}) reduces  to 
 \begin{equation}
     \ddot{\phi}+3H\dot{\phi}\approx0 \implies\eta=-\frac{\ddot{\phi}}{H\dot{\phi}} \approx3.
 \end{equation}
This temporary increment in the value of the second slow-roll parameter also supports the existence of the USR phase during inflation, which later causes a rapid growth in the amplitude of the curvature power spectra that provides the necessary seed for the formation of the PBHs. 
 \par
\begin{figure}
    \centering
\includegraphics[width=0.50\textwidth]{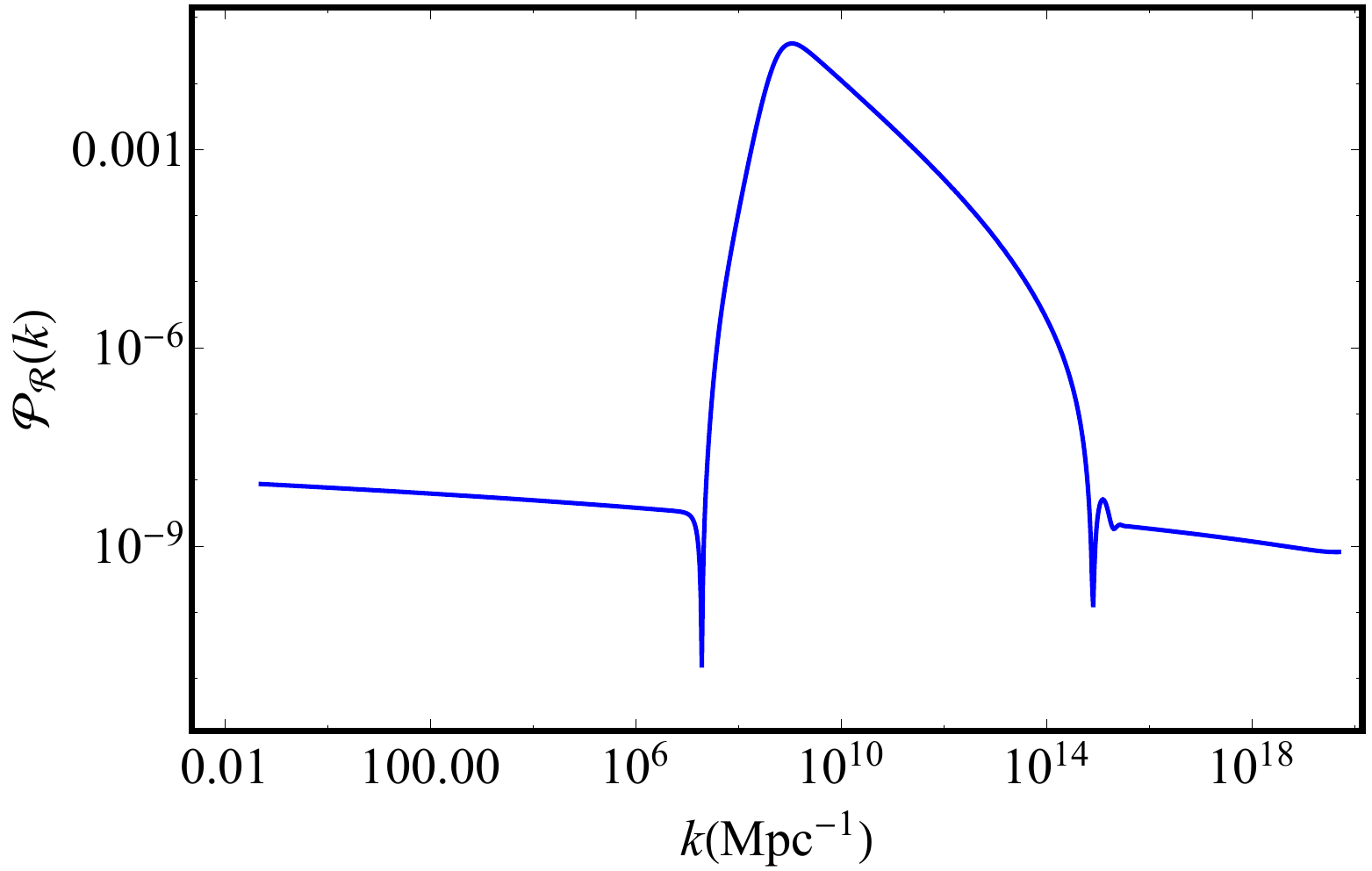}
    \caption{Diagram of amplitude curvature power spectrum $P_{\mathcal{R}}(k)$ from the Mukhanov-Sasaki equation at  $k_{pk} \sim 1.12712 \times10^9$ Mpc$^{-1}$}
    \label{fig:powerspectra}
\end{figure}

\begin{figure}
    \centering
\includegraphics[width=0.55
\textwidth]{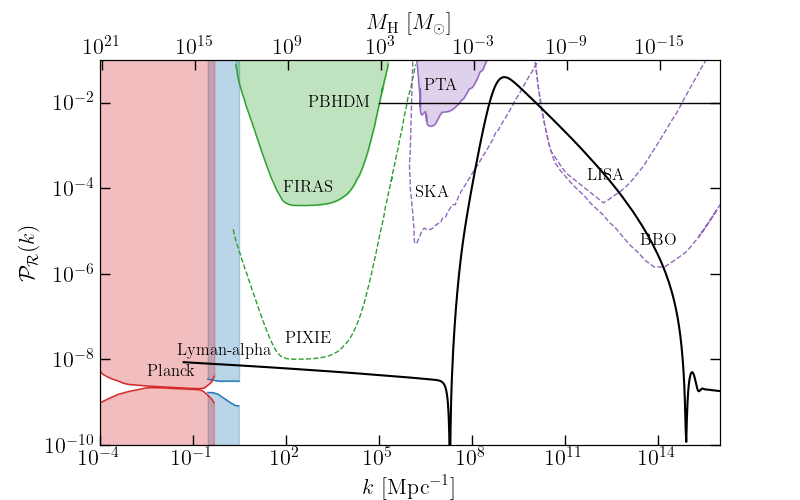}
    \caption{ Diagram illustrating the  evolution of the amplitude of the power spectra $P_{\mathcal{R}}(k)$ during the SR-USR-SR phase in the background of  constraints on the primordial power spectrum  derived from both  ongoing (CMB, Lyman $\alpha$-forest, PTA) and future experiments (SKA, LISA, BBO) \cite{green2021primordial}}
    \label{fig:powerspectra2}
\end{figure}

 We have plotted the amplitude of the scalar power spectrum in momentum space using Eq. (\ref{eq:mukhanovpowerspectrum}) in Fig. \ref{fig:powerspectra}. Interestingly, it is clear from Fig. \ref{fig:powerspectra} that the value of the amplitude of $ P_\mathcal{R}(k)\sim 10^{-9}$ agrees well with the COBE-normalized value for large scales (small $k$-value) during slow-roll inflation. In Fig. \ref{fig:powerspectra}, the resulting scalar curvature power spectrum $P_{\mathcal{R}}(k)$ also shows enhancement at small scales during the USR phase. In particular, at these scales, it is found that the amplitude of the power spectrum $P_{\mathcal{R}}(k)$ reaches a value of $\sim 10^{-2}$ from $10^{-9}$ at $k_{pk} \sim 1.12712 \times10^9$ Mpc$^{-1}$. After completion of the USR phase, when the inflaton again transits into the SR phase, the power spectra again join hands with the COBE-normalized amplitude. This substantial amplification in the power spectrum during the early inflationary phase, where the inflaton undergoes a phase of USR, causes the gravitational collapse signaling the possibility of PBH formation during the radiation-dominated era when the horizon scale matches the scale of perturbation \cite{Novikov1979Primordial}. In Fig. \ref{fig:powerspectra2}, we have plotted the evolution of the same amplitude of the curvature power spectra in the background of various constraints derived from ongoing and future observations \cite{green2021primordial}. It is observed from Fig. \ref{fig:powerspectra2} power spectrum extends into the forecasted sensitivity constraint of the Square Kilometer Array (SKA).   
 In the following section, we determine various characteristics of the PBHs, including their mass, evaporation time, Hawking temperature, and the resulting present-day abundance, which is quantified by the PBH fraction \( f_{\rm PBH} \).

\section{\label{sec:fpbh}Formation of PBHs and Abundance}
 These enhanced modes remain super-Hubble after their generation and, immediately after the horizon re-entry ($k=aH$) during the radiation-dominated era, act as the seed for the production of PBHs through gravitational collapse \cite{Mahbub:2019uhl}. The expressions for the PBH mass associated with the enhanced modes and their formation fraction and present-day abundance can be determined through the Press-Schechter formalism \cite{1974ApJ...187..425P, DELUCA2020135550, Young:2024jsu}. Following \cite{merchand2026exploring, Mahbub:2019uhl}, we first outline a general prescription for the PBH formation fraction, where we mainly focus on two key parameters: the mass, and their abundance, namely $f_{PBH}$ for the PBHs from the Press-Schechter formalism.

\par
\textbf{Mass of the PBH:} The re-entry of modes is deeply connected with the mass of the PBHs that are formed via collapse of perturbations. As expected, the collapsing region does not put all the horizon mass into the formation of PBHs; the PBH mass is a fraction and is proportional to the horizon mass, given by \cite{Mahbub:2019uhl} 
{\small  \begin{equation}
\begin{split}
    M_\mathrm{PBH}(k)&=\gamma M_\mathrm{hor}=\frac{4\pi}{3} \gamma \rho H_k^{-3}|_{k=aH}\\&= 1.6 \times10^{18}g\left(\frac{\gamma}{0.2}\right)\left(\frac{g_*(T_k)}{106.75}\right)^{-1/6}\left(\frac{k_{pk}}{5.5\times10^{13} \mathrm{Mpc}^{-1}}\right)^{-2}
\end{split}
\label{eq:massPBH}
\end{equation} }
where  $\gamma$ is the collapse efficiency factor $=0.2$  for radiation domination and $g_* (T_{rad})=106.75$ is the effective number of degrees of freedom at PBH formation deep in the radiation epoch \cite{saikawa2018primordial}.

\begin{figure}
    \centering
\includegraphics[width=0.55\textwidth]{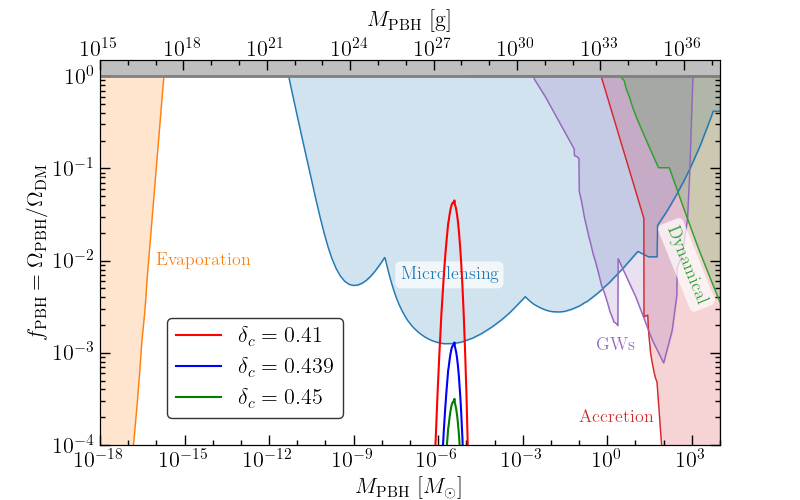}
    \caption{Diagram of the PBH abundance $f_{PBH}$ v/s the mass of the PBH for the stringy inflaton potential. The results are obtained for different values of density contrast $\delta_c$, illustrating the sensitivity of the $\delta_c$ for the corresponding mass. }
    \label{fig:fpbh}
\end{figure}
\par \textbf{ Press-Schechter formalism:} The Press-Schechter formalism \cite{1974ApJ...187..425P}, which was originally developed to describe the formation and abundance of large-scale structures such as galaxies and clusters of galaxies, can also be implemented in the sector of PBHs. The Press-Schechter framework generally quantifies a fraction of mass from the matter density of regions that form into objects due to gravitational collapse whose density contrast exceeds a critical value \cite{escriva2026beyond, PhysRevD.106.123519}.  The Press-Schechter framework is applicable in the context of PBHs, as upon re-entering the horizon the modes that exceed the critical density $\delta_c$ are assumed to be responsible for the formation of the PBHs. In this context, the threshold value for PBH formation depends on the shape of the density perturbation. For the broad peaks of $\delta$, the value tends towards $0.41$, which is the minimum threshold for PBH formation \cite{musco2019threshold}. In contrast, for the sharp peaks, the value reaches $0.66$, which is regarded as the maximum threshold \cite{musco2019threshold}. We can estimate the fraction of PBHs formed with a mass $M(K)$ over total radiation  energy at the time of formation using  \cite{Mahbub:2019uhl, merchand2026exploring,ballesteros2018primordial}
\begin{equation} 
\begin{split}
\beta(M(k))=\int_{\delta_c}^{\infty}
\frac{1}{\sqrt{2\pi}\sigma(M(k))}
\exp\left[-\frac{\delta^2}{2\sigma^2(M)}\right]d\delta,
\\=\frac{1}{2}
{\rm erfc}\left[
\frac{\delta_c}{\sqrt{2}\sigma(M(k))}
\right].
\label{eq:betaequation}
\end{split}
\end{equation}
Here, $\sigma^2 (M(k))$ is the mean square of coarse-grained density perturbations on a smoothing scale $R=k^{-1}$ and $\delta$ is the density contrast. The perturbation which having $\delta$ greater than the critical value $\delta_c$ survives the radiation pressure and collapses gravitationally \cite{Mahbub:2019uhl}. The mass variance is defined as \cite{Mahbub:2019uhl, merchand2026exploring}
\begin{equation}
\sigma^2(M(k))=\frac{16}{81}
\int_0^\infty d\ln q
\left({q}{R}\right)^4P_{\mathcal R}(q)
W^2\left({q}{R}\right)
\label{eq:sigma}
\end{equation}

with 
\begin{equation}
W\left({q}{R}\right)
=\exp\left[-\frac{1}{2}\left({q}{R}\right)^2\right],
\label{eq:W}
\end{equation} is a Gaussian window function which smooths the density perturbation over a particular physical length scale. The modes of inflation predict the power spectrum in Fourier space, while the abundance of the PBH is calculated in real space, which is convolved by the window function W \cite{PhysRevD.97.103528}. It is noteworthy that the formation of PBH depends on a particular scale rather than a particular Fourier mode. Thus, the window function W determines how strongly each mode contributes to the formation of the PBH over a particular scale. Since the PBH behaves as matter and the background of the universe is radiation in the radiation-dominated era, the energy density of PBHs increases as the universe goes through expansion.   It is therefore useful to express the PBH abundance in terms of the present-day fraction of dark matter contained in PBHs, defined as
\begin{equation}
\begin{split}
f_{PBH}=\frac{\Omega_{PBH}(M)}{\Omega_{DM}}  =\\\frac{\beta(M)}{8\times10^{-16}}
\left(\frac{\gamma}{0.2}\right)^{3/2}
\left(\frac{106.75}{g_(T_f)}\right)^{1/4}
\left(\frac{M}{10^{18}{\rm g}}\right)^{-1/2}
\label{eq:fbh}
\end{split}
\end{equation}
Using Eqs. (\ref{eq:betaequation}), (\ref{eq:sigma}), and (\ref{eq:W}) in Eq. (\ref{eq:fbh}), for the modified string inflaton potential in (\ref{Vmodified}), we have determined the abundance of the generated PBHs for different values of the density contrast $\delta_c$ and plotted them in Fig. \ref{fig:fpbh}. It is evident from Fig. \ref{fig:fpbh} that the abundance of PBHs is highly sensitive to the density contrast since the formation probability $\beta (M(k))$ depends exponentially on the density threshold. The predicted PBHs from the string inflaton potential are concentrated around $10^{-6} M_\odot$, where the constraints on the dark matter fraction arise from microlensing bounds \cite{green2021primordial, PhysRevD.99.083503}. The microlensing bounds (shaded region) represent the upper limit of the $f_{PBH}$ from the absence of lensing events, confirmed by various observations such as EROS, OGLE, Subaru/HSC, and MACHO. In particular, the OGLE  observation tightens the upper bounds on the Earth-Jupiter mass PBHs, i.e.,  $M_{PBH}={10^{-6}-10^{-3}} M_\odot$ \cite{PhysRevD.99.083503}. In this context, it should be noted that a theoretical prediction lying below the bounds, i.e., the shaded region, is compatible with observations, whereas a prediction lying inside the shaded region is in tension with present observations. We have plotted the abundance for three threshold values of density contrast $\delta_c$; in particular, for $\delta_c=0.439$ it just touches the edge of the microlensing constraints \cite{green2021primordial}. For the limiting density constraint, the stringy inflaton model produces $0.12\%$ of the total DM. This $\delta_c$ essentially limits the lowest threshold value of density contrast for contemporary microlensing observations for the stringy inflaton potential. Further decrease in the value of the density contrast essentially leads the abundance to lie inside the constraints region, which is not compatible with microlensing observations \cite{green2021primordial}.  
\par Having determined the abundance and the mass of the PBHs, another important aspect is their evaporation time and Hawking temperature. It is noteworthy that these quantities are directly related to the survival and contribution to the total energy budget of the universe. If the PBHs evaporate only by thermal radiation, emitting all the allowed particles permitted by the Standard Model, then their evaporation time and Hawking temperature are given by    
 \cite{Cline_1997}
    \begin{equation}
    t_\mathrm{eva}=4.89\times 10^{2} \left(\frac{M_\mathrm{PBH}}{10^{9}}\right)^3 ms
    \label{eq:evaporationtime}
\end{equation} and \cite{PhysRevD.110.063024}
\begin{equation}
    T_\mathrm{H}=1.06\times\frac{10^{13}\mathrm{g}}{M_\mathrm{PBH}( g)}\quad\mathrm{GeV}.
    \label{eq:hawkingtemperture}
\end{equation}  

\par
 It is evident from Fig. \ref{fig:powerspectra} that the amplitude of the  curvature power spectra enhanced at \(k_{pk} \sim 1.12 \times 10^{9}\) Mpc\(^{-1}\). Using this $k$-value in Eq. (\ref{eq:massPBH}), gives a mass of \(3.80 \times 10^{27}\) grams, which is approximately \(1.91 \times 10^{-6} M_{\odot}\) for the generated PBHs. The mass derived for these PBHs generated from the stringy inflaton potential corresponds to a nearly planet-scale mass but is sufficiently below that of ordinary stellar black holes. Utilizing the calculated mass in (\ref{eq:evaporationtime}) and (\ref{eq:hawkingtemperture}), we have calculated the evaporation time ($t_{eva}$) and the Hawking temperature ($T_{H}$) for the PBHs created from the modified stringy inflaton potential. The evaporation time and the Hawking temperature are $\sim2.70 \times 10^{55}$ sec $ \sim8.56\times 10^{38}$ Gyr  and $\sim2.78 \times 10^{-15}$ GeV, respectively. The extremely large evaporation time, which is many orders of magnitude greater than the present age of the universe, i.e., $\sim$ 13.79 Gyr \cite{riess2019large,carmeli2006cosmic, NANDI2025170155}, indicates that the PBHs produced from the stringy model can survive in the present epoch.   This also makes them a suitable candidate for DM. In this context, it should be noted that PBHs with a mass less than $5\times 10^ {14} $ g cannot exist due to complete evaporation by Hawking radiation \cite{Mosbech:2022lfg}, which are categorized as light PBHs and are not suitable as DM candidates.  
 
\section{\label{sec:conclusion} Conclusions}
In the present paper, we study the formation of PBHs from a stringy inflaton potential modified with a Gaussian bump. To consider the PBH formation that happened in the early universe, the string theoretic model can be a good consideration. Here, a stringy potential which is capable of producing successful inflation \cite{Let:2023dtb,sarkar2025study} is modified by introducing a Gaussian bump. This modification is inspired by a localized deformation of one of the K\"ahler moduli that induces a USR phase in the potential. To determine the amplitude of the curvature power spectrum, we have numerically solved the Mukhanov-Sasaki equation beyond the standard slow-roll approximation using the Bunch-Davies initial condition. When the inflaton transitions from the SR to the USR phase, a subsequent amplification in the curvature power spectrum, i.e., $10^{7}$ times larger than CMB observations, is achieved, which is the necessary seed for PBH formation. This amplified curvature perturbation, when re-entering the horizon during radiation domination, influences the density fluctuations of radiation to create gravitational collapses, ultimately leading to PBH formation.\par
 The numerical analysis of various quantities along with the curvature power spectra originating from the stringy inflaton potential reveals a clear departure from the standard slow-roll scenario near the region of the Gaussian bump.  In Fig. \ref{fig:slow-roll}, it is shown that inclusion of the Gaussian bump flattens the variation of the scalar field even more than standard slow-roll, indicating that the inflaton moves more slowly. Figures \ref{fig:Slow-roll parameter} and \ref{fig:sSlow-roll parameterd} show that the first and second slow-roll parameters also deviate from the standard slow-roll scenario, indicating a phase of USR. It is found that the curvature power spectra show a significant enhancement of $10^{-2}$ near $k\sim 1.12712\times 10^{9}$ $Mpc^{-1}$, which is suitable for PBH formation, as shown in Fig. \ref{fig:powerspectra}. We have also plotted the enhancement of the curvature power spectra against the bounds derived from various ongoing and future experiments in Fig. \ref{fig:powerspectra2}. It is found that the enhancement in the curvature power spectra lies in the sensitivity region of the SKA observations, suggesting a possibility that the scalar-induced gravity waves may be probed by future observations.   \par
 Using the Press-Schechter formalism, we have also calculated the abundance of the PBHs and examined the viability of our model against the available bounds for different values of the critical density contrast. The enhancement of the amplitude curvature power spectra occur near $k\sim 1.12712\times 10^{9}$ $Mpc^{-1}$ which corresponds to a solar mass $\sim 10^{-6} M_\odot$. For the corresponding mass range, we have found that if the density contrast is lower than $0.439$, the calculated abundance of dark matter is in tension with the current microlensing observations. Further, we have also calculated the evaporation time $\sim2.70 \times 10^{55} Sec$, and the Hawking temperature $\sim2.78\times10^{-15} GeV$ for the PBHs. The evaporation time is far longer than the current age of the universe. This also suggests that the PBHs formed from the stringy inflaton model are stable against the Hawking temperature over a cosmological timescale. In conclusion, it can be stated that the deformation of one of the k\"{a}hler moduli in the moduli space leads to a correction in the original potential in field space, inducing a USR phase during slow-roll inflation that produces long-lived Earth-Jupiter mass PBHs that may serve as a plausible DM candidate. 
 \section*{Acknowledgements}
  C.S. thanks Dr. Rafid Mahbub
for useful discussions and clarifications on the numerical solution of the Mukhanov-Sasaki equation regarding PBH formation.
 \section*{Data availability statement}
Data sharing does not apply to this article since no datasets were created or analyzed during this research.
%%%%%%%%%%%%%%%%%%%%%%%%%%%%%%%%%%%%%%%%%%%%%%%%%%%%%%%%%%%%%%%%%%%%%%%%%%%%%%%%%%%%%%%%%%%%
\section*{Code availability statement}
Code/Software sharing does not apply to this article, as the present study did not include the generation or analysis of any code or software.
%%%%%%%%%%%%%%%%%%%%%%%%%%%%%%%%%%%%%%%%%%%%%%%%%%%%%%%%%%%%%%%%%%%%%%%%%%%%%%%%%%%%%%%%%%%%
\section*{Conflict of interest}
The authors declare that they have no conflicts of interest to disclose.
\bibliography{biblio}% Produces the bibliography via BibTeX.

\end{document}